\documentclass[10pt]{article}

\usepackage{PRIMEarxiv}
\usepackage{authblk}
\usepackage[utf8]{inputenc} 
\usepackage[T1]{fontenc}    
\usepackage{hyperref}       
\usepackage{url}            
\usepackage{booktabs}       
\usepackage{amsfonts}       
\usepackage{nicefrac}       
\usepackage{microtype}      
\usepackage{lipsum}
\usepackage{fancyhdr}       
\usepackage{graphicx}       
\usepackage{xspace}
\graphicspath{{media/}}     
\date{}
\newcommand{\sys}{DroidBot-GPT\xspace} 

\title{DroidBot-GPT: GPT-powered UI Automation for Android
}

\author[1]{Hao Wen}
\affil[1]{Institute for AI Industry Research (AIR), Tsinghua University}

\author[2]{Hongming Wang}
\affil[2]{Beijing University of Posts and Telecommunications}

\author[2]{Jiaxuan Liu}

\author[1]{Yuanchun Li}

\begin{document}
\maketitle


\begin{abstract}
This paper introduces \sys, a tool that utilizes GPT-like large language models (LLMs) to automate the interactions with Android mobile applications. 
Given a natural language description of a desired task, \sys can automatically generate and execute actions that navigate the app to complete the task.
It works by translating the app GUI state information and the available actions on the smartphone screen to natural language prompts and asking the LLM to make a choice of actions.
Since the LLM is typically trained on a large amount of data including the how-to manuals of diverse software applications, it has the ability to make reasonable choices of actions based on the provided information.
We evaluate \sys with a self-created dataset that contains 33 tasks collected from 17 Android applications spanning 10 categories. It can successfully complete 39.39\% of the tasks, and the average partial completion progress is about 66.76\%. Given the fact that our method is fully unsupervised (no modification required from both the app and the LLM), we believe there is great potential to enhance the automation performance with better app development paradigms and/or custom model training.

\end{abstract}

\section{Introduction}

\begin{figure}
    \centering
    \includegraphics[width=16.4cm]{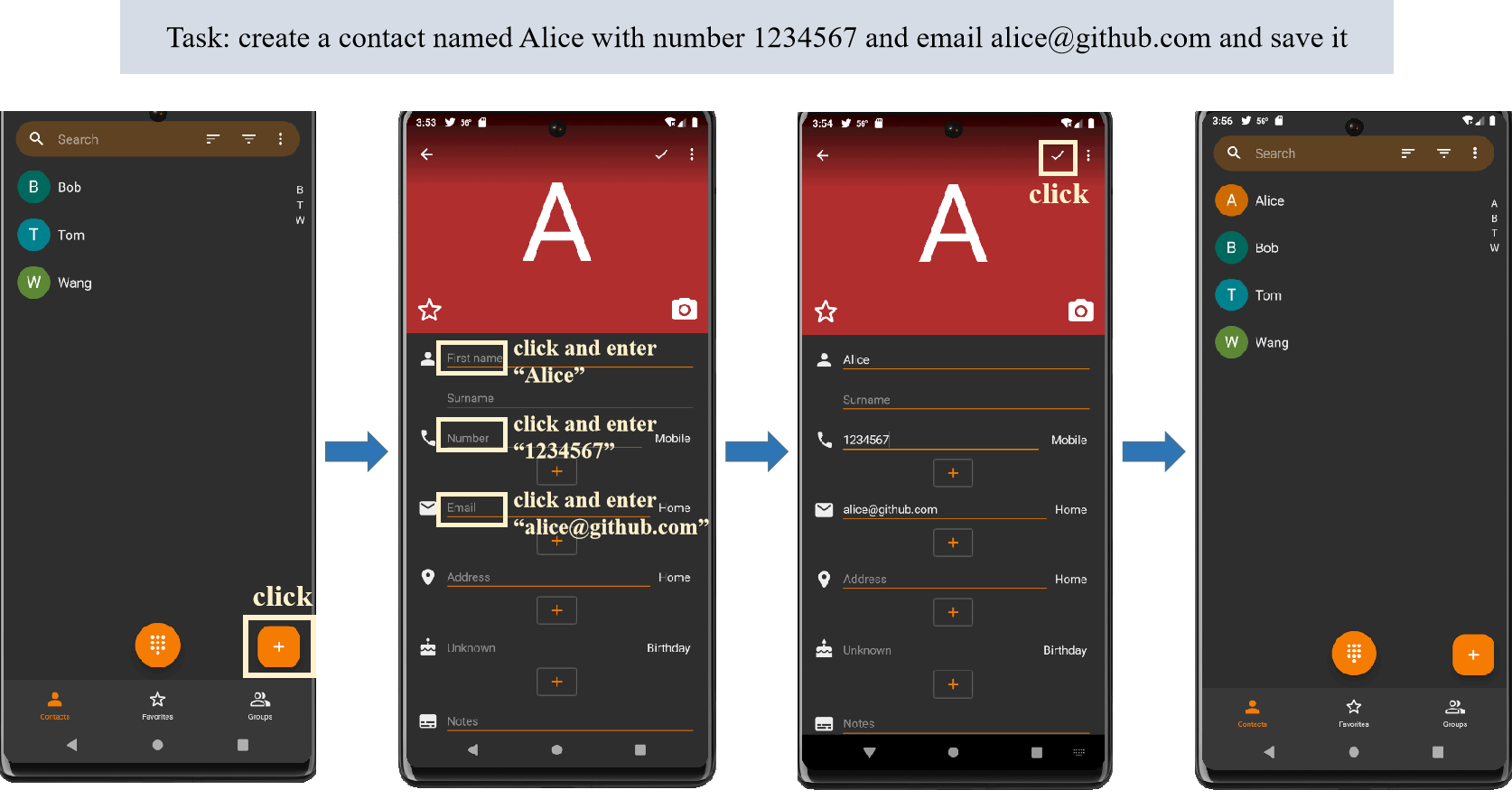}
    \caption{An illustration of task automation with \sys. The task is to use the ``Simple Contacts Pro'' app to complete the task ``create a contact named Alice with number 1234567 and email alice@github.com''.}
    \label{fig:sys-example}
\end{figure}


UI automation is a fundamental technology used in various applications such as robotic process automation, software testing, and intelligent personal assistants.
Among them, natural language-grounded automation is especially attractive due to the improved user experience. Existing products, such as Siri, Amazon Alexa, and Google Assistant, typically follow a two-step approach. Firstly, they employ NLP techniques to extract the intent of a task and input parameters from a natural language request. Then, they execute the corresponding actions by connecting an API to a specific command. Recent research approaches have introduced end-to-end techniques for fulfilling users' requests \cite{mapping_mobile, motif, seq2act}. These methods are becoming increasingly popular due to their flexibility and adaptability to user needs, as well as their ability to understand complex language and provide accurate, clear, and natural responses.
However, how to support the automation of diverse applications with little or no development effort remains a challenging problem.

Recent advances in large language models (LLM) have significantly enhanced the performance of numerous natural language processing (NLP) tasks, paving the way for novel approaches to automating tasks that were once exclusively carried out by humans. For example, ChatGPT \cite{chatgpt} and other similar pre-trained generative models have shown to be effective on a variety of tasks in the financial domain \cite{bloomberggpt}, the healthcare industry \cite{chatgpthealth}, and even robot control \cite{robotgrounding}. The success of LLMs in these fields inspires us to think, whether and how will large language models benefit the natural language-grounded automation of mobile applications.

This paper focuses on LLM-powered Android application automation in which natural language descriptions will be interpreted into a series of actions on a smartphone directly (Figure \ref{fig:sys-example}). 
For each graphical user interface, we leverage DroidBot \cite{droidbot} to extract structured information, translate it into a natural language prompt, query an LLM with the prompt to generate the next action, and execute the actions on the smartphone.
By properly designing the prompt, \sys can clearly interpret the task and UI information to the LLM, and let the LLM return a correct and concise action.

In order to evaluate our system, we manually design 33 tasks to be executed on 17 Android applications and analyze the completion rate. Our system can complete most of the straightforward tasks such as \textit{``Add a new task with the content of completing the homework on ToDoList app''}. Besides, our system can even finish some complex tasks needing multiple GUI jumps and some prior knowledge like \textit{``search the first holiday of Bavaria in 2023''}.

We summarize our contributions below:
\begin{itemize}
    \item We present \sys, a GUI automation tool for Android that utilizes a large language model (LLM) to interact with mobile applications. To the best of our knowledge, we are the first to investigate how pre-trained language models can be applied to app automation without app and model modifications.
    \item We introduce how to automatically generate natural language to describe tasks, states, and actions in mobile applications. By grounding the LLM, we are able to identify the appropriate action sequence for these tasks. We contend that this technique presents an interesting opportunity for using large language models in decision-making and robotic process automation.
\end{itemize}


\section{Method}
\begin{figure*}
    \centering
    \includegraphics[width=14cm]{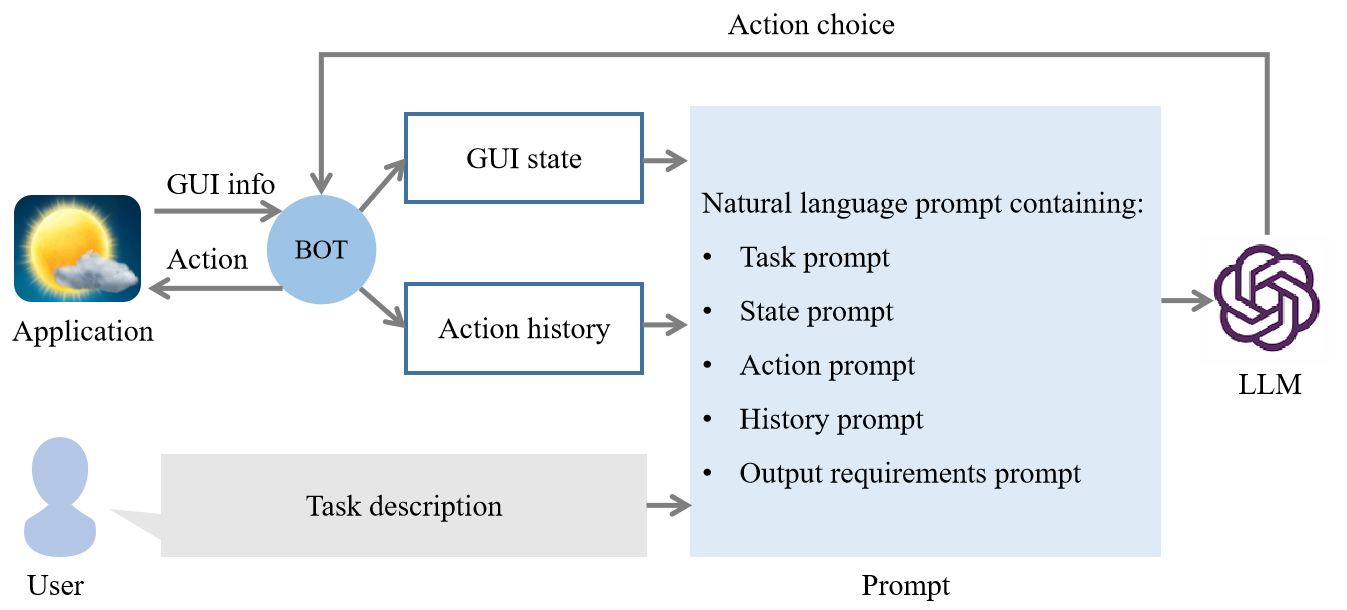}
    \caption{The workflow of \sys.}
    \label{fig:system_architecture}
\end{figure*}
Figure \ref{fig:system_architecture} illustrates the overall architecture of our system. Given an Android application and a task described by the user, \sys first fetches the state of the app and describes it in natural language. Then it combines the state information, the action history, and the task into a prompt, and sends it to ChatGPT \cite{chatgpt}. ChatGPT generates the sends back the proper action, and \sys sends operations to be executed on the phone.


\subsection{GUI description}

The process of translating a graphical user interface into a form that a large language model can handle is a challenging task in our system. A potential method would be to input the structured, tree-form description of the GUI interface directly. However, this method tends to present thousands of words, which frequently contain useless information and surpasses the length limit. As a result, we propose an alternative approach, which involves converting the GUI interface into a familiar natural language sentence that can be more readily processed by GPT models.

\begin{figure}
    \centering
    \includegraphics[width=10cm]{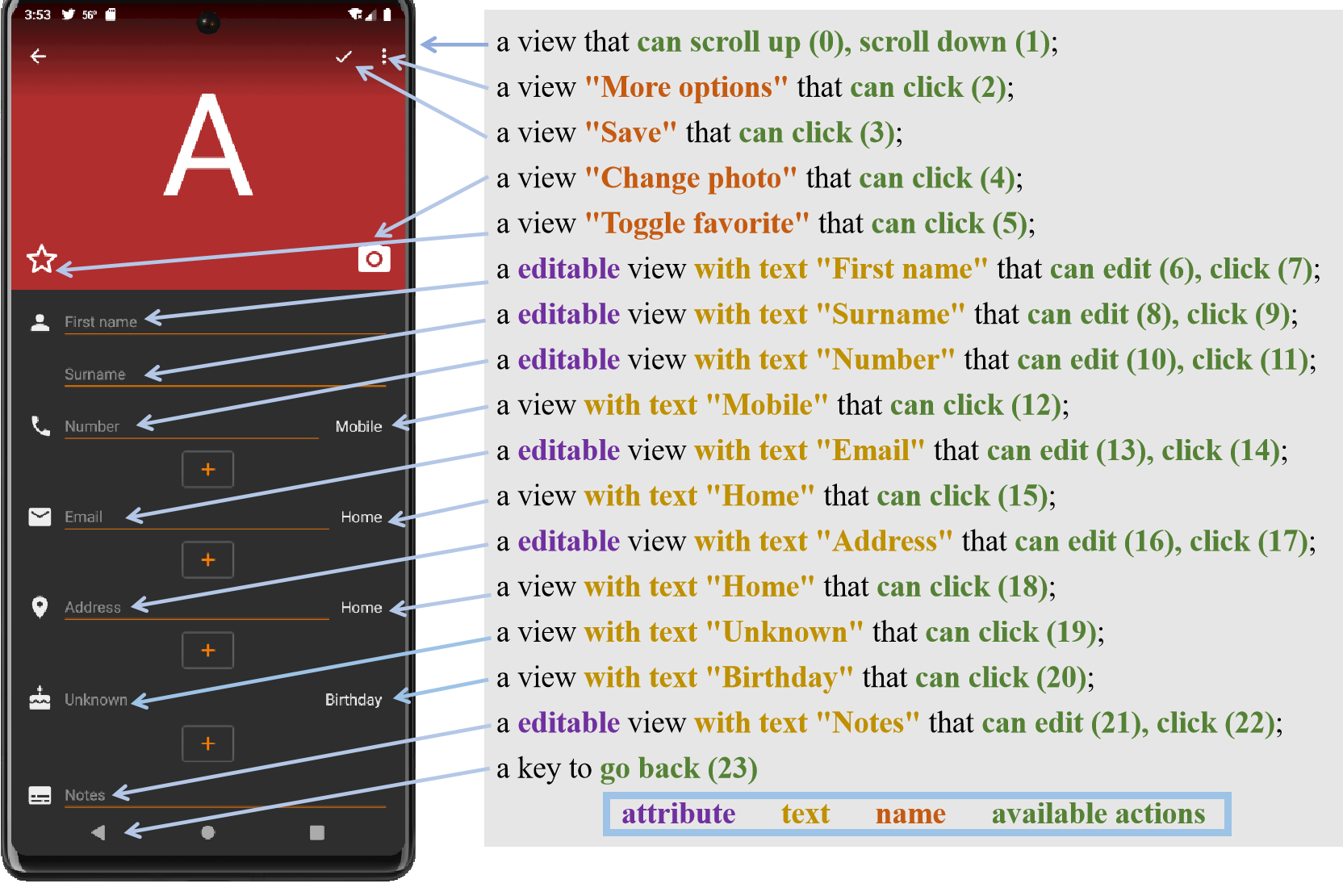}
    \caption{An example of interpreting GUI elements into prompts. For an element, the attribute includes ``editable'', ``clicked'', and ``selected''. The available actions include ``click'', ``long click'', ``check'', ``edit'', ``scroll up'', and ``scroll down''. }
    \label{fig:element_attributes}
\end{figure}

Given a graphical user interface, we first extract all the user-visible elements and check their properties. For each element, we generate a prompt \textit{``A view <name> that can ...''} succeeded by all the property prompts shown in Figure \ref{fig:element_attributes}. Then, we combine all the elements with a previous text \textit{``The current state has the following UI views and corresponding actions, with action id in parentheses''}. We can transform a UI tree into a natural language sentence using the rules above.



\subsection{Action space}

The actions available on mobile devices can be categorized into choosing and editing. The choices mainly include click, scroll, check, and so on, which have been shown in Figure \ref{fig:element_attributes}. In the prompts that describe the elements in a GUI, we add a number behind it, which we utilize ChatGPT to choose from. For example, suppose that a GUI description prompt contains \textit{``a view `Sort by' that can click (0);}''. When we input the prompt to ChatGPT and get a response \textit{``0''}, then we execute the action \textit{``click the button {`Sort by'}''}.

Editing means typing sentences in text boxes. Users sometimes have to input a username, password, or sentence, which can not be encoded into choices. Therefore, we design a two-step solution. When the LLM model chooses to edit in a text box, \sys sends another prompt ended with \textit{``What should I enter to the view with the text `<text content>'? Just return the text and nothing else.''}, after which \sys types in the response. One challenge is that the responses from ChatGPT may not meet the requirement. For example, it may return a sentence instead of a choice, in which case \sys will extract the first choice it contains. 

\subsection{Prompt synthesis}

\begin{figure*}
    \centering
    \includegraphics[width=16.4cm]{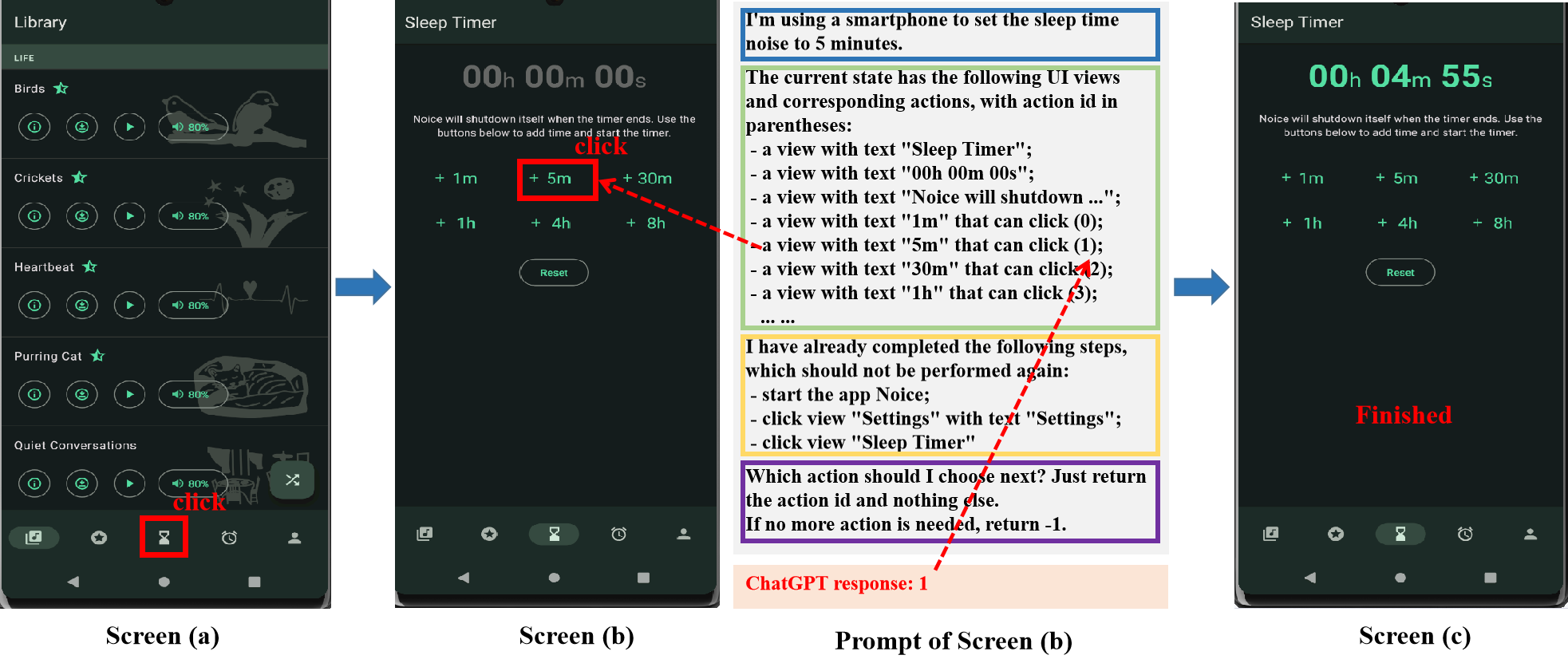}
    \caption{An example of the prompt synthesized by \sys. We select a music player application capable of playing white noise for a specified duration. The task is \textit{``set the sleep time noise to 5 minutes''}. The proper action sequence includes \textit{``start the app {`Noice'}''}, \textit{``click view `Sleep Timer' (Screen (a))''}, and \textit{``click view with text `5m' (Screen (b))''}. To provide a clear understanding, we are representing the task, GUI elements, action history, and output requirement in blue, green, yellow, and purple colors respectively. }
    \label{fig:prompt_example}
\end{figure*}

Besides GUI descriptions and the action space, the prompt should also include a history action sequence to avoid repetition. Therefore, the prompt consists of the task, the GUI elements with the action choices they provide, the action history, and the requirement whether the output should be a single choice or a sentence that is to be typed in. An example of the prompt is shown in Figure \ref{fig:prompt_example}.

\section{Evaluation}

We evaluate our system on 17 broadly used Android applications downloaded from F-Droid \cite{Fdroid}, which is an application market focused on Free Open Source Software (FOSS) on the Android platform. We design 1$\sim$3 tasks for each application, containing 2$\sim$13 steps on graphical user interfaces. We show an application example, tasks designed based on it, and their proper action sequences in Table \ref{tab:task_example}.

An issue with prompt generation is that a singular element in a GUI could have an excessive amount of text, causing the prompt to be too lengthy for large language models to process, leading to the exclusion of crucial information or later elements. As a result, we have established a maximum text length of 20 words based on our experiments.

\begin{table}
	\caption{An example of tasks designed for World Weather (an application that reports current weather and forecasts around the world).}
	\centering
	\begin{tabular}{ll}
		\toprule
		 Task   & Actions   \\
		\midrule
    Check the current temperature of London              & start the app World Weather -> \\
    & click view with text "London" \\
    \midrule
	    Add Beijing and check its current temperature    & start the app World Weather -> \\
     & click view 'Add city' ->\\
     & click view with text "city, country" -> \\
     & enter "Beijing China" into view with text "city, country" -> \\
     & click view 'search' -> \\
     & click view with text "Beijing, CN  (39.91,..." -> \\
     & click view with text "Current  weather" \\
     \midrule
     Close the wind direction display                    &  start the app World Weather -> \\
     & click view 'Extras' ->\\
     & click view with text "Settings" -> \\
     & click view with text "Wind direction display" \\     
		\bottomrule
	\end{tabular}
	\label{tab:task_example}
\end{table} 

\subsection{Results}
In order to evaluate the model on the tasks not fully completed, we record the actions \sys generates, which are compared with the proper action sequence. Suppose the proper action sequence is $A_{true}=\{a_1, a_2, ... a_n\}$, and \sys-generated actions are $A_{pred}=\{a_1, a_2, ... a_i, a_{i+1}', a_{i+2}', ... a_{m}'\}$, where $\{a_{i+1}, a_{i+2}, ... a_{n}\} \neq \{a_{i+1}', a_{i+2}', ... a_{m}'\}$, the completion progress of $A_{pred}$ is $P_{pred}=\frac{i}{n}$. 

\sys fully completes 13 out of 33 tasks, and the average completion progress of all tasks is 66.76\%. Table \ref{tab:task_completion} shows the average completion progress for tasks of different complexity levels, where the straightforward tasks with 2$\sim$3 steps are easier to be completed. This is intuitive because the language model has not been fine-tuned to understand the interface relations. 

Table \ref{tab:category_completion} shows the completion rate of tasks from different categories. The average completion progress of applications from ``Record'' and ``Life'' are relatively low, because these tasks often include a lot of typing actions, and our system can miss the right place to input due to the ambiguous GUI element relationship.

Compared to the previous works that are designed for a specific range of applications or actions \cite{worldofbits, glider, mapping_mobile}, \sys does not require further GUI training data and can efficiently generalize tasks across a larger array of categories, thanks to the utilization of large language models. Furthermore, our system obviates the need for in-depth instruction guidance, as opposed to previous works like \cite{mapping_mobile}. Our system can successfully complete complex tasks with minimal instructions, thereby achieving a high degree of automation.

\begin{table}
	\caption{The completion rate and total number of tasks at each complexity level.}
	\centering
	\begin{tabular}{lccc}
		\toprule
    Task complexity  & Number & Average completion progress & Fully completion rate  \\
		\midrule
    2$\sim$3 steps   & 10     & 73.33\%   & 60.00\% \\
    4$\sim$5 steps   & 13      & 63.46\%  & 38.46\% \\
    6$\sim$13 steps  & 10      & 66.97\%  & 20.00\% \\
    Total            & 33      & 66.76\%  & 39.39\% \\
		\bottomrule
	\end{tabular}
	\label{tab:task_completion}
\end{table} 

\begin{table}
	\caption{The completion rate and total number of tasks for each category.}
	\centering
	\begin{tabular}{lccc}
		\toprule
            Task category           & Number  & Average completion progress  & Fully completion rate \\
		\midrule
            System                  & 3       & 76.67\%          & 33.33\% \\
            Record                  & 5       & 41.19\%          & 20.00\% \\
            Weather                 & 3       & 70.24\%          & 33.33\% \\
            Financing               & 3       & 92.86\%          & 33.33\% \\
            Sports and Health       & 3       & 66.67\%          & 33.33\% \\
            Science and Education   & 3       & 75.00\%          & 66.67\% \\ 
            Calls and texts         & 3       & 84.62\%          & 33.33\% \\
            Music                   & 2       & 62.50\%          & 50.00\% \\
            Life                    & 2       & 50.00\%          & 50.00\% \\
            Tools                   & 6       & 70.67\%          & 50.00\% \\
		\bottomrule
	\end{tabular}
	\label{tab:category_completion}
\end{table} 

\subsection{Limitations}

Although \sys demonstrated a high level of automation and strong generalization in certain experiments, it is important to address the cases where it failed. The failures can be categorized into three types: early termination, redundant operation, and GPT's response not meeting the requirements. Our investigation found unnamed GUI elements, obscure GUI relationships, and the uncertainty of actions to be the primary reasons for the failures.

\begin{figure*}
    \centering
    \includegraphics[width=13cm]{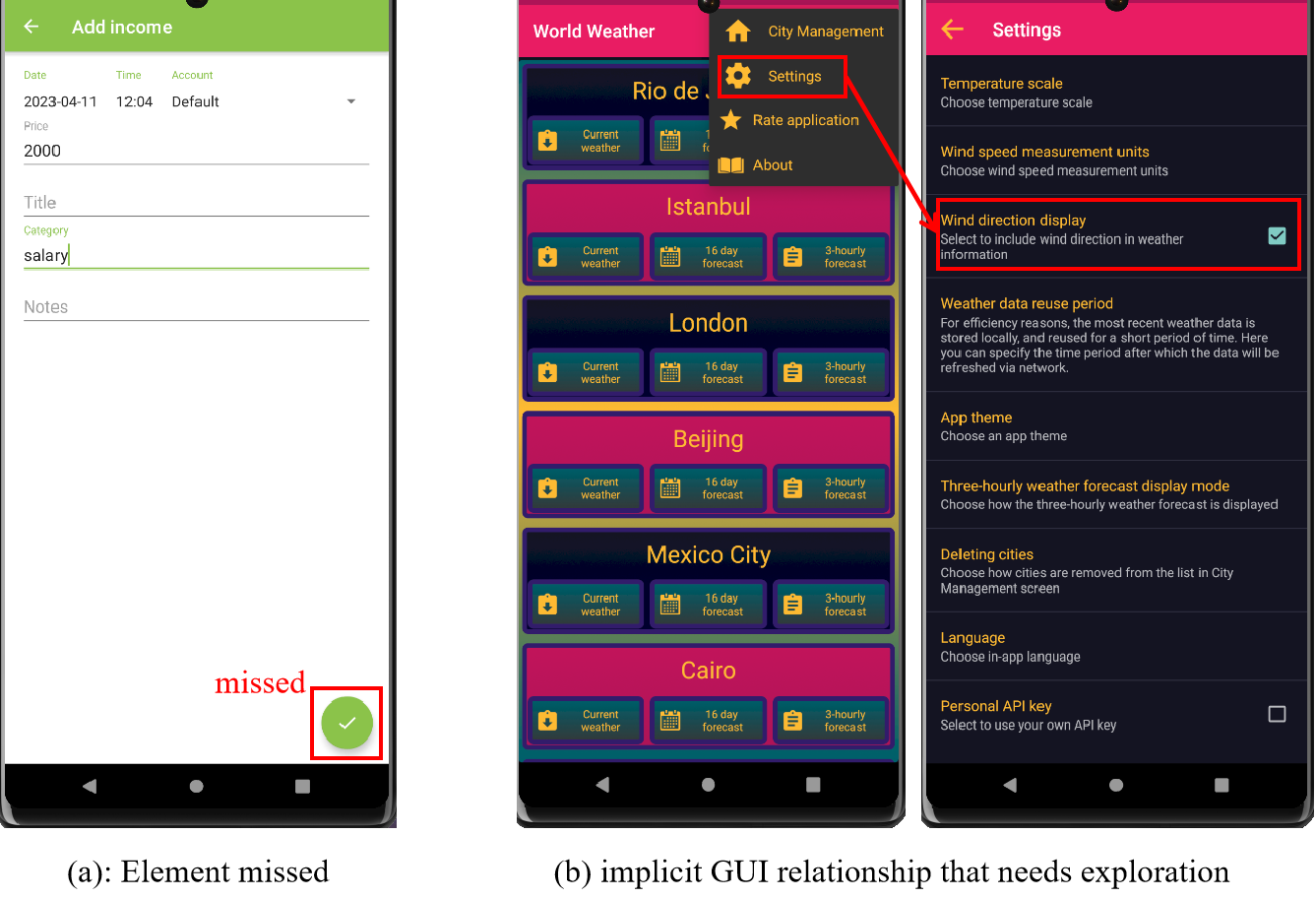}
    \caption{We present two examples of failure cases. Figure (a) displays an unnamed GUI element that is not recognized by DroidBot. The unnamed check mark button in the bottom right, which is required to save the operations, is absent in the action space due to its absence in the GUI tree, ultimately resulting in failure. Figure (b) shows an instance where we requested the "World Weather" app to disable the wind direction display. The implicit relationship between the "Wind direction display" button and the "Settings" button requires further exploration; hence, the LLM cannot provide a suitable option in Figure (b).}
    \label{fig:failure}
\end{figure*}
\textbf{Unnamed GUI elements.} Certain GUI elements lack text and are challenging to identify, especially when they are crucial to completing tasks, such as checkmark buttons or search boxes (e.g. Figure \ref{fig:failure}(a)). In these cases, \sys may become confused, resulting in the early termination of the program or incorrect button clicks.

\textbf{Obscure GUI relationships.} In certain instances, the relationship between GUI elements and screens can be ambiguous. As illustrated in Figure \ref{fig:failure}(b), when attempting to complete tasks that require exploring a particular application, \sys may execute redundant or incorrect operations.

\textbf{The uncertainty of actions.} The response generated by ChatGPT is probability-based, resulting in occasional instability of the system's actions. For instance, when repeating a task that \sys can complete multiple times, the order of operations may vary slightly, which can potentially result in redundancy or even failure.

These failure cases suggest limitations and challenges that need to be addressed in future work.
We will try to fix the aforementioned problems by fine-tuning LLM and using machine-learning methods to recognize unnamed elements.
\section{Related work}

\textbf{GUI test generation} is to automatically create test scripts that verify the functionality of the graphical user interface, aiming to reduce the manual effort required to test GUI applications. There are various types of strategies for GUI test generation, including random \cite{monkey, dynodroid, Sapienz}, model-based \cite{droidbot, test_study, droidmate}, search-based \cite{search_based},  record/replay testing \cite{capture/replay}, and learning-based \cite{humanoid}.
Our system utilizes DroidBot \cite{droidbot} to generate input for Android applications after receiving the instructions from the LLM.

\textbf{Digital task automation} mainly includes web and mobile application task automation. Many researchers have attempted to train AI agents to interact with the web given a task description with natural language processing and reinforcement techniques \cite{worldofbits,mapping_web,glider}. WebGPT \cite{webgpt} fine-tunes GPT-3 to answer long-form questions using a text-based web-browsing environment. Inspired by these works, \sys also uses an agent to interact with applications and designs a similar action space including clicking and editing.

Li et al. \cite{mapping_mobile} introduce the new task of generating mobile user interface actions guided by natural language instruction, and creates three new datasets correspondingly. MOTIF \cite{motif_eccv} creates a large dataset for mobile app task automation and task feasibility prediction. Lexi \cite{lexi} uses a pre-trained vision and language model to extract information from mobile and desktop UI to interact with it. Screen Correspondence \cite{screen_correspondence} investigates a machine learning method to reason about UIs. Unlike these works, \sys introduces a large language model-powered method to interact with UIs, which has greater flexibility and scalability, 
\section{Conclusion}
This paper introduces \sys, the first Android app automator guided by a large language model. We demonstrate how to translate a GUI and its corresponding action space into prompts and queries to the language model, enabling it to perform tasks specified by the user. We evaluate the performance on a dataset consisting of tasks of varying complexity and categories. We also identify the limitations of our system and propose potential solutions to overcome them. Our ultimate goal is to enable the user experience where large language models can assist individuals in decision-making and task completion.



\bibliographystyle{unsrt}  
\bibliography{references}

\end{document}